\documentstyle[preprint,oldlfont,aps,doublespace]{revtex}
\begin{document}
\title{Design of proteins with selected thermal properties}

\author{M. P. Morrissey$\dag$ and E. I. Shakhnovich $\ddag$\\ $\dag$
Department of Applied Physics, Harvard University\\ $\ddag$ Department
of Chemistry, Harvard University} \maketitle

\pagebreak

\begin{abstract}
We propose a new and effective means for designing stable and
fast-folding polypeptide sequences using a cumulant expansion of the
molecular partition function.  This method is unique in that $T_{Z}$,
the ``cumulant design temperature'' entered as a parameter in the
design process, is predicted also to be the optimal folding
temperature.  The method was tested using monte-carlo folding
simulations of the designed sequences, at various folding temperatures
$T_{F}$.  (Folding simulations were run on a cubic lattice for
computational convenience, but the design process itself is
lattice-independent.)  Simulations confirmed that, over a wide range
of $T_{Z}$, all designed sequences folded rapidly when $T_{F} \approx
T_{Z}$.  Additionally, highly thermostable model proteins were created
simply by designing with high $T_{Z}$.  The mechanism proposed in
these studies provides a plausible pathway for the evolutionary
design of biologically active proteins, which {\em must} fold and
remain stable within a relatively narrow range of temperatures.

\end{abstract}
\newpage

The functionality of a protein stems from the molecule's ability to
assume one unique shape, the ``native conformation,'' out of an
untenable sea of possible conformations.  Nature seems to have {\em
designed} protein sequences capable of folding into specific
conformations \cite{LEVIN,GO,GSW,PRLF}.  The problem of sequence
design is the ``inverse folding'' problem: given a 3-dimensional
target structure, predict a sequence of amino acids which will
spontaneously fold to that structure, and remain stable with respect
to that structure.

A recent systematic approach to the sequence design problem utilizes
the idea of stochastic optimization in sequence space \cite{PNAS}: one
should design sequences with a large energy gap between the native
conformation and the set of misfolded, denatured states. The simplest
implementation was based on the approximation that the free
energy of the denatured state does not depend on sequence (only on
amino acid composition), while the energy of the native state is
sequence-dependent.  Subsequent improvements to this algorithm
accounted in a crude way for the dependence of the energy of unfolded
conformations on sequence, thus relaxing the condition of fixed amino
acid composition \cite{LOCAL}.  The same approach with minor technical
modifications was used in the recent work \cite{DK}. No folding
simulations were made to test the sequences designed in this work,
however, making it impossible to evaluate the design algorithm of
\cite{DK} as well as the applicability of the heteropolymer model used
there.

The serious shortcoming of all previous sequence design methods is the
conspicuous lack of reference to a folding temperature $T_{F}$.
Stability and foldability are, however, sensitive functions of $T_{F}$
\cite{TRAPS,SO2,LOCAL}.  Ideally, one would like to design sequences
which are foldable within a preimposed, biologically relevant
temperature range.  Since the stability of the native state is
determined by the difference in free energy between the unfolded state
and the native state, a realistic treatment of the temperature
dependence of the free energy of the denatured state is paramount to any
reasonable attempt of ``rational'' protein design.

In this work, we report a method of rational sequence design which
takes a target structure {\em and a desired optimal folding
temperature $T_{Z}$}, and generates a sequence which is
thermodynamically stable with respect to the target structure {\em at
a folding temperature $T_{F} \approx T_{Z}$}.  (See Figure 1.)
\marginpar{Fig.1} The method, which we call the ``cumulant design
method,'' is based on a mean-field high temperature expansion of the
single-molecule partition function, which allows us to estimate the
free energy of the denatured state explicitly.

The energy of a model protein can be evaluated as
\begin{equation}
\label{energy}
E = \sum_{i=1}^{N} \sum_{j<i} B(\xi_{i},\xi_{j})\Delta_{ij}
\end{equation}
where $\Delta_{ij}=1$ if monomers $i$ and $j$ are within some
specified distance range in three-dimensions, and $0$ otherwise.
$\xi_{k}$ is the identity of the amino acid at position k, and $B$ is
the parameter set matrix, a symmetric energy matrix representing the
pairwise attraction (or repulsion) of the various amino acids.  We use
the phenomenological parameter set of Miyazawa and Jernigan \cite{MJ}.

\bigskip
The probability of finding a chain in any given state during the
folding process is given by the canonical ensemble:
\begin{equation}
P(\{\vec{x}\}, \{\sigma\}, T) =
\frac{ \prod_{i=1}^{N-1}g(\vec{x}_{i+1}-\vec{x}_{i})
\exp(-E(\{\vec{x}\}, \{\sigma\})/ T)} {Z(\{\sigma\},T)}
\end{equation}
where the set $\{\vec{x}\}$ represents the positions of the N
monomers, $\{\sigma\}$ represents the amino acid sequence, and $g$
explicitly represents the constraints imposed by the chain.\cite{LGK}
(We absorb Boltzmann's constant $\kappa_{B}$ into $T$.)  The
conformation-space partition function is:
\begin{equation}
\label{conf_Z}
Z(\{\sigma\},T) = \sum_{\{\vec{x}_{1} \ldots
\vec{x}_{N}\}} \prod_{i=1}^{N-1}g(\vec{x}_{i+1}-\vec{x}_{i})
\exp(-\frac{E(\{\vec{x}\},\{\sigma\})}{T})
\end{equation}
It is through the partition function $Z$ that properties of the
unfolded state affect the stability of the native state.

We can rewrite the partition function as an intergral over a
continuous density of states:
\begin{equation}
\label{continuous_Z}
Z(T)= \gamma^{N-1} \int_{-\infty}^{\infty} e^{-E/T_{Z}} \rho (E) dE
\end{equation}
where $\gamma^{N-1}$ is the total number of possible conformations and
$\rho(E)$ is the normalized density of energy states.  The
{\em cumulants} ${{\langle c_{n} \rangle}}$ of $\rho(E)$ are defined by
\begin{equation}
\label{cumulant_def}
\log \int_{-\infty}^{\infty} e^{-i t E} \rho (E) dE
= \sum_{n=1}^{\infty} \frac{(-i t)^{n}}{n!} {{\langle c_{n} \rangle}}.
\end{equation}
Solving for $\rho(E)$ and substituting into (\ref{continuous_Z}) yields
\begin{equation}
\label{cumulant_expansion}
Z(T) = \frac{\gamma^{N-1}}{2 \pi}
\exp \left( \sum_{n=1}^{\infty} \frac{(-1)^{n}}{n! T^{n}} {{\langle c_{n} \rangle}} \right).
\end{equation}
This is a high-temperature expansion of $Z(T)$.

It is useful that the cumulants of independent random variables add
linearly.  Thus if we know the cumulants of the energy probability
distribution for a {\em single contact}, the energy cumulants of a
system with $N_{C}$ independent, identically distributed contacts is
simply
\begin{equation}
\label{single_total}
\mbox{$\{ {{\langle c_{1} \rangle}} \ldots {{\langle c_{n} \rangle}}\}$} =
\mbox{$\{N_{C}{{\langle c_{1} \rangle}}_{s}
\ldots N_{C}{{\langle c_{n} \rangle}}_{s} \}$}.
\end{equation}

Of course, (\ref{cumulant_expansion}) is useful only if we can
estimate the cumulants of $E$.  We make this estimate in the
mean-field, by first calculating the set of simple moments
${{\langle\mu_{n} \rangle}}_{s}$ of the {\em single contact} energy $E_{s}$:
\begin{equation}
\label{moments}
{{\langle \mu_{n} \rangle}}_{s} = \frac{\sum_{i,j=1}^{N} P_{ij}
[B(\xi_{i},\xi_{j})] ^{n}} {\sum_{i,j=1}^{N} P_{ij}}
\end{equation}
In our mean-field approximation, monomers $i$ and $j$ are assumed,
during random motion, to interact with a probability $P_{ij}$,
determined by the chain connectivity.  Since the correlation length in
globules is small \cite{LGK}, a good estimate of $P_{ij}$ for the
globular, denatured state is the uniform distribution over {\em all
allowed} contacts \cite{CHANDILL2,CONSTR}.  (On a cubic lattice, this
excludes contacts for which $i-j$ is even.)

We utilize the high temperature expansion (\ref{cumulant_expansion})
and the noted properties of cumulants to present the cumulant
algorithm for sequence design:
\medskip

1.  Begin with a target structure and a random sequence.  The target
structure will remain fixed, as the optimization occurs in sequence
space.

2.  Compute the first $n$ moments of the single-contact energy
(eq. \ref{moments}).  

3.  Compute the first $n$ single-contact cumulants in terms of the
first $n$ moments, using a known recursion relation \cite{NORMAL}.

4.  Utilize (\ref{single_total}) to approximate the cumulants of the
total molecular energy.  We used the maximally compact $N_{C}$, which
was shown computationally to be the optimal choice.

5.  Estimate the partition function using (\ref{cumulant_expansion}).
Note that the temperature used in (\ref{cumulant_expansion}) is our
cumulant design temperature $T_{Z}$, which is predicted to be the
optimal folding temperature.

6.  Calculate the fitness parameter, the canonical probability of
native state occupancy.  For convenience, we actually maximize the
quantity $P_{Z}$ which is monotonic with $P$:
\begin{equation}
\label{PZ}
P_{Z}(T) = -E_{N} - \sum_{n=1}^{\infty} \frac{(-1)^{n}}{n! T^{n-1}} <c_{n}>
\end{equation}

7.  Make a random mutation in sequence space.

8.  Compute $P_{Z}$ for the mutated sequence, and accept or reject the
mutation according to the Metroplis algorithm.  One must use a
convenient ``sequence space selective  temperature'' $T_{sel}$
\cite{PNAS} to run the simulation; this is unrelated to the cumulant
design temperature $T_{Z}$.

9.  Continue the monte-carlo search until equilibrium is reached, and 
$P_{Z}$ is (approximately) maximized.

\bigskip
The lack of an accurate ``energy field'' for real proteins currently
prevents us from designing foldable proteins in the laboratory
setting, so we must test our design hypothesis within the context of
computational models.  For simplicity we chose a standard cubic
lattice model.  Under such a model, each monomer is represented as a
vertex on a cubic lattice, and two monomers ``interact'' if they are
lattice neighbors but do not occupy successive positions on the chain.
It is important to note that the cumulant design method itself is not
beholden to any particular model; the cubic lattice is used merely to
{\em test} the method.

We designed a number of sequences for different target backbones and
with various design temperatures $T_{Z}$, and tested the thermodynamic
and kinetic properties (stability and folding times, respectively) of
these sequences.  For our target (native) structures, we used five
random, maximally compact 36-mer backbones.  We used several
structures in order to make sure that any properties noted are not
artifacts of a particular native conformation.

Ten sequences were designed for each of five cumulant design
temperatures ($T_{Z} = 0.20, 0.28, 0.36, 0.44, 0.75$), on each of the
five random backbones, for a total of 250 designed sequences.  The
cumulant expansion was cut off at fifteen terms, enough to infer
convergence of the series.  Sequences were designed using the
aforementioned Miyazawa-Jernigan parameter set \cite{MJ},
and were ``optimal'' in the sense that they maximized $P_{Z}(T)$
within the approximations used.

We ran $10$ folding simulations, at various temperatures, on each of
the designed sequences.  The runs were $1,000,000$ monte-carlo (MC)
steps in length.  The folding algorithm was that of Hilhorst
\cite{DEUTCH}, utilizing three standard moves: crankshaft, corner
flip, and tail moves.  One MC step consisted of one randomly chosen
(sterically possible) move, whether or not that move was accepted
according to the Metropolis criterion.

Figure 2 demonstrates that the optimal folding temperature for a
sequence is strongly correlated with the the cumulant design
temperature $T_{Z}$ for that sequence.  
\marginpar{Fig.2}
This is true over a factor of
two in absolute temperature; within this range, the minimal folding
time occurs for $T_{F} \approx T_{Z}$.  The relationship breaks down
at high $T_{Z}$, for reasons that will be discussed subsequently.

The designed sequences were also tested for stability.  Starting from
the folded state, sequences were subjected to folding simulations of
$10,000,000$ MC steps.  Figure 3 shows clearly that sequences
designed with high $T_{Z}$ were more thermostable than sequences
designed with low $T_{Z}$.  (We acknowledge that longer runs would be
necessary to confirm the stability data at very low $T_{F}$.)

The sequences designed by the cumulant algorithm fold at least as
rapidly as those designed for maximal gap (for comparison see
e.g. \cite{TRAPS,LOCAL}); the cumulant method could even design
fast-folding 64-mers (median folding time 2.1 million MC steps).  Most
importantly, the cumulant design temperature $T_{Z}$ proved a good
predictor of the optimal folding temperature, allowing us to design
sequences with specified (optimal) $T_{F}$.

Of particular interest is the newfound ability to design thermostable
sequences. To this end it is instructive to study which features of
sequences are responsible for optimal behavior at high folding
temperatures $T_{F}$.  One can keep only two terms in the cumulant
expansion to approximate the optimized quantity given by (\ref{PZ}):
\begin{equation}
\label{approximate_P}
P_Z \approx -E_N + N_C(B_0 - \frac{\sigma^2}{2T})
\end{equation}
(a similar expression for the partition function of a disordered
heteropolymer was obtained in \cite{BIOCH,Stepanow}) where $B_0$ is
the average interaction energy and $\sigma^2$ is its variance.  It is
clear from (\ref{approximate_P}) that at low temperature, sequences
with minimal dispersion should be selected, perhaps at the expense of
compromising low target state energy $E_N$, while at higher $T$ this
factor is less important and what is mostly optimized is $E_N$. Figure
4 shows that this is precisely what occurs.

Although our algorithm is effective over a factor of two to three in
absolute temperature, it is {\em not} generally possible to design
sequences which are stable at temperatures outside of this range.  It
was effectively impossible to design sequences with $T_{Z} < 0.20$,
due to divergence of the high-temperature expansion
(\ref{cumulant_expansion}).  At high $T_{Z}$, design efficacy eroded
due to limits on the interaction energies; no molecule can be stable
at temperatures higher than the characteristic $T$ of its strongest
interactions. Additionally, our assumption of a compact denatured
state becomes progressively less accurate as $T$ increases.

\medskip
Any one protein can indeed fold only within a specific thermal
environment.  One example is the enzyme ribonucleotide
reductase\cite{Sando:1973}.  The version of this enzyme present in the
mesophilic bacterium {\em Lactobacillus leichmannii} is maximally
active at $49^{\circ} C$, while its counterpart in the thermophile
{\em Thermus X-1} demonstrates maximal activity at about $70^{\circ}
C$.  Importantly, the activity curves are narrow: at $49^{\circ}$, the
thermophilic enzyme is less that 30\% active, and at $70^{\circ}$, the
mesophilic homologue is completely inactive.  Although it might be
impossible to design a single protein which folds at all habitable
temperatures, nature has been able to give each organism the proteins
it needs to thrive in its own thermal environment.

Our cumulant design method seems to give a reasonable estimate of
the molecular partition function: we are able to design model proteins
which are stable (and foldable!) at a given temperature.  We thus have
a useful model for the thermodynamic properties of real proteins.

\medskip
This work was supported by the Hertz Foundation for the Applied
Physical Sciences and the Packard Foundation.  
We are grateful for useful discussions with Sharad
Ramanathan and Victor Abkevich.

\section{Figures}
{\bf Figure 1}: Schematic of the sequence design process.  The goal is to
design a sequence which folds to a specific conformation, at a
pre-specified design temperature $T_{Z}$.

{\bf Figure 2}: Median folding time vs. folding temperature (36-mer).
250 sequences were designed for five randomly-chosen, maximally
compact backbones, at five cumulant design temperatures $T_{Z}$.  Each
sequence was folded 10 times at each of several temperatures $T_{F}$,
and folding times were sorted by $T_{Z}$ and $T_{F}$.  For $T_{Z} \leq
0.5$, optimal folding occurs near $T_{F} = T_{Z}$, as predicted.
(Folding runs were cut off at 1 million steps.)

{\bf Figure 3}: Stability vs. folding temperature (36-mer).  The same 250
sequences designed for Figure 2 were tested for denaturation.
Starting from the native state, the sequences were subjected
10,000,000 monte-carlo steps.  Here, stability is defined as the
percentage of MC time spent with $\geq 39$ of the $40$ native contacts
intact.

{\bf Figure 4}: Energy of the target conformation (left axis) and dispersion
of interaction energies (right axis) for sequences designed
at different $T_Z$

\bibliographystyle{unsrt}

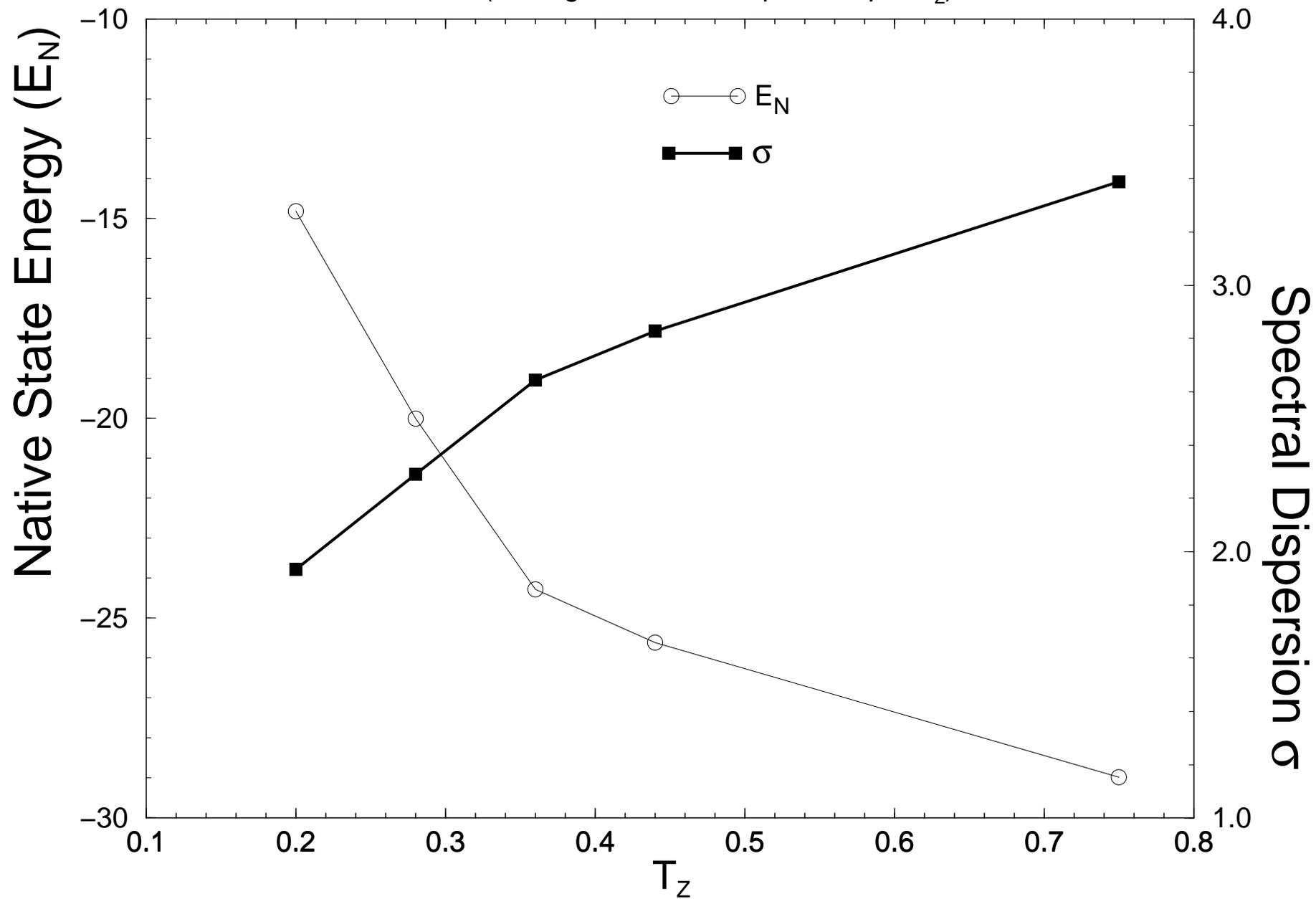
\end{document}